\documentclass[conference]{IEEEtran}
\IEEEoverridecommandlockouts
\usepackage{cite}
\usepackage{amsmath,amssymb,amsfonts}
\usepackage{algorithmic}
\usepackage{graphicx}
\usepackage{textcomp}
\usepackage{xcolor}
\def\BibTeX{{\rm B\kern-.05em{\sc i\kern-.025em b}\kern-.08em
    T\kern-.1667em\lower.7ex\hbox{E}\kern-.125emX}}

\graphicspath{{figures/}}
\usepackage[colorlinks,
            linkcolor=blue,
            anchorcolor=blue,
            urlcolor=blue,
            citecolor=blue
            ]{hyperref}

\usepackage{bbding}
\usepackage{algorithmic}
\usepackage{algorithm}
\makeatletter
\newcommand{\removelatexerror}{\let\@latex@error\@gobble}
\makeatother
\usepackage{graphicx}
\usepackage{multirow}
\usepackage{booktabs}

\usepackage{authblk}
\usepackage{relsize}

\usepackage{tikz}
\usetikzlibrary{shapes.geometric, arrows}
\usetikzlibrary{arrows.meta,positioning,fit,backgrounds,shadows.blur,calc}

\begin{document}

\title{\textsc{ForgeDAN}: An Evolutionary Framework for Jailbreaking Aligned Large Language Models\\
\thanks{$^{\dagger}$ The authors contributed equally to the work.}
}

\author{
    Siyang Cheng$^{\dagger1,2}$, Gaotian Liu$^{\dagger1,2}$, Rui Mei$^{*1,2,3}$, Yilin Wang$^{1,4}$, Kejia Zhang$^{1,5}$, Kaishuo Wei$^{6}$\\Yuqi Yu$^{7}$,  Weiping Wen$^{3}$, Xiaojie Wu$^{1,2}$, Junhua Liu$^{2}$\\
    \smaller
    $^{1}$\textit{iFLYTEK Security Laboratory, Hefei, China}\\
    $^{2}$\textit{Anhui SparkShield Intelligent Technology Co., Ltd., Hefei, China}\\
    $^{3}$\textit{Peking University, Beijing, China}\\
    $^{4}$\textit{School of Automation, University of Electronic Science and Technology of China, Chengdu, China}\\
    $^{5}$\textit{Northwest University, Xi'an, China}\\
    $^{6}$\textit{University of New South Wales, Sydney, Australia}\\
    $^{7}$\textit{National Computer Network Emergency Response Technical Team/Coordination Center of China
, Beijing, China}\\
    $^{*}$Corresponding author: \textbf{ruimei@pku.edu.cn}
}

\maketitle

\begin{abstract}

The rapid adoption of large language models (LLMs) has brought both transformative applications and new security risks, including jailbreak attacks that bypass alignment safeguards to elicit harmful outputs. Existing automated jailbreak generation approaches e.g. AutoDAN, suffer from limited mutation diversity, shallow fitness evaluation, and fragile keyword-based detection. To address these limitations, we propose \textsc{ForgeDAN}, a novel evolutionary framework for generating semantically coherent and highly effective adversarial prompts against aligned LLMs. First, \textsc{ForgeDAN} introduces multi-strategy textual perturbations across \textit{character, word, and sentence-level} operations to enhance attack diversity; then we employ interpretable semantic fitness evaluation based on a text similarity model to guide the evolutionary process toward semantically relevant and harmful outputs; finally, \textsc{ForgeDAN} integrates dual-dimensional jailbreak judgment, leveraging an LLM-based classifier to jointly assess model compliance and output harmfulness, thereby reducing false positives and improving detection effectiveness. Our evaluation demonstrates \textsc{ForgeDAN} achieves high jailbreaking success rates while maintaining naturalness and stealth, outperforming existing SOTA solutions.

\end{abstract}

\begin{IEEEkeywords}
jailbreak attack, adversarial prompt generation, large language models (LLMs), evolutionary algorithm, AI safety
\end{IEEEkeywords}

\section{Introduction}
\label{Introduction}

In recent years, the rapid evolution of artificial intelligence (AI), particularly in the field of large-scale generative models, has ushered in a new era of Artificial Intelligence Generated Content (AIGC). Among these, large language models (LLMs) e.g. ChatGPT, Gemini, and Claude have become emblematic of this transformation. Their unprecedented capability to understand, generate, and interact in natural language and other modalities has not only reshaped the landscape of human-computer interaction but also accelerated the integration of AI into both personal and industrial scenarios such as personal writing assistance and email summarization to critical applications including healthcare consultation, legal reasoning, scientific discovery, and educational support, LLMs have demonstrated remarkable adaptability and versatility\cite{ zhang2025stair, kasri2025vulnerability, hu2024llm4mdg}.

Despite their transformative potential, the rapid deployment of LLMs has also raised significant concerns regarding safety, security, and controllability. While these models are designed to follow human instructions, their probabilistic nature and reliance on vast training corpora make them vulnerable to generating outputs that deviate from ethical or legal standards. Existing studies have shown that LLMs may inadvertently produce violent narratives, explicit sexual content, misinformation, politically sensitive discourse, or discriminatory expressions, depending on the prompts they receive\cite{augenstein2024factuality, sakib2024risks, kim2025assessing, sun2024topic, lin2022truthfulqa}. Furthermore, their generative capacity can lead to emergent behaviors that are difficult to predict, raising the possibility of "loss of control" where the model generates unsafe or undesirable outputs despite alignment efforts. To mitigate these risks \cite{owasp_2023_llm_top_10, sun2023mitigating}, researchers have developed alignment techniques such as Supervised Fine-Tuning (SFT)\cite{bianchi2024safety} and Reinforcement Learning from Human Feedback (RLHF)\cite{ji2025safe}, which constrain model behavior by teaching refusal strategies and ethical boundaries. Although effective to some extent, these methods only reduce, rather than eliminate, the risks inherent in large-scale generative systems.

Nevertheless, alignment safeguards are not unbreakable. Increasing evidence shows that adversarially constructed prompts—carefully designed sequences of text that exploit the model’s instruction-following tendencies—can bypass these protections and induce harmful or policy-violating outputs. A notorious example is the Do-Anything-Now (DAN) prompt series, which encourages the model to abandon its aligned behavior and operate in an unconstrained mode, thereby enabling the generation of dangerous, unethical, or prohibited content\cite{shen2024anything}. Systematic investigations have further revealed hundreds of jailbreak prompts in the wild, demonstrating that diverse prompting strategies, including role-playing, obfuscation, and indirect questioning, can repeatedly circumvent content filters\cite{liu2023jailbreaking}. These findings underscore a critical gap between the intended safety alignment objectives of LLMs and their actual behavior when confronted with adversarial manipulation, highlighting the necessity of developing more rigorous and resilient evaluation frameworks to probe and fortify model security.

Current jailbreak attack strategies fall into two main categories, i.e. manually crafted prompts and automated adversarial methods. Manual approaches e.g. DAN-based role-playing and reverse induction are creative and effective but rely heavily on human ingenuity and struggle to scale or adapt. Automated methods like gradient-guided attacks e.g. Greedy Coordinate Gradient (GCG) \cite{zou2023universal} generate adversarial strings programmatically, but often result in nonsensical or garbled prompts that are susceptible to simple detection techniques, such as perplexity-based filtering. AutoDAN-HGA (AutoDAN with Hierarchical Genetic Algorithm) \cite{liu2024autodan} improves on this by using a hierarchical genetic algorithm to automatically evolve more natural, stealthy jailbreak prompts. Still, it has limitations: low diversity due to reliance on single-path mutations, semantic insensitivity in fitness evaluation (e.g., token-level Jaccard similarity), and brittle jailbreak detection via keyword matching that may yield false positives or overlook partial responses \cite{ganguli2022red}.

To address these limitations, we propose \textsc{ForgeDAN}, a novel evolutionary jailbreak framework that enhances and extends the AutoDAN-like approaches. We introduce multi-strategy textual perturbations—including character, word and sentence-level techniques—to generate diverse and semantically coherent adversarial prompts. For fitness evaluation, we also employs a semantic similarity model that compares model outputs before and after mutations, delivering more interpretable and meaningful assessments. Furthermore, \textsc{ForgeDAN} leverages a LLM-based classifier to robustly determine whether the model refused to answer or produced harmful content, thus improving detection accuracy and reducing both false positives and false negatives.

The contributions of this paper are as follows:  
\begin{itemize}  
  \item Multi-strategy text perturbation approach is integrated, encompassing character, word, and sentence-level mutations, to enhance diversity in adversarial prompt generation.
  \item We introduce a semantic similarity fitness metric model, enabling more interpretable and effective selection of strong prompt candidates.  
  \item An optimized LLM-based semantic classifier is deployed for robust and accurate detection of harmful or refusal responses, improving the reliability of jailbreak evaluation.  
  \item We design and implement the prototype of \textsc{ForgeDAN}, an evolutionary jailbreak mechanism that produces semantically fluent and highly effective DAN-style prompts.
\end{itemize}

\section{Related Work}
\label{Related Work}

Jailbreak attacks on LLMs have drawn increasing attention. Existing studies relied on manual prompts, later evolving into automated generation methods and output detection techniques. Together, these works form the foundation for advancing jailbreak research on aligned models.

\subsection{Manual Jailbreak and Red Teaming Techniques}
The traditional jailbreak attacks on LLMs relied heavily on human creativity, where adversarial prompts were carefully crafted to exploit model vulnerabilities \cite{wallace2022discovering}. Several notable pieces of work systematically analyzed in-the-wild jailbreak prompts. For example, Liu et al. collected 78 verified jailbreak prompts and proposed a taxonomy of three main categories—camouflage, attention diversion, and privilege escalation—covering ten specific prompting patterns\cite{liu2023jailbreaking}. They further constructed a dataset of 3,120 jailbreak issues aligned with OpenAI’s usage policies \cite{openai2025usage}, targeting eight prohibited scenarios and evaluating vulnerabilities in ChatGPT-3.5 and 4.0. Shen et al. provided another landmark contribution by analyzing the infamous "Do-Anything-Now (DAN)" prompts, showing how role-playing and persona-shifting strategies enabled LLMs to bypass safety alignment and produce prohibited outputs\cite{shen2024anything}. Other studies introduced manipulations such as suffix injection and refusal suppression, revealing how subtle modifications could drastically reduce the effectiveness of refusal mechanisms\cite{zhou2024don}.  

Manual red teaming has also been widely employed as an evaluation approach. In this setting, human experts actively probe LLMs through interactive testing, attempting to induce harmful outputs that circumvent alignment constraints. Such practices are valuable since they emulate realistic adversarial behaviors and yield qualitative insights into model vulnerabilities. Nevertheless, significant limitations remain: manually crafted jailbreaks are labor-intensive, lack scalability across diverse models or tasks, and often lose effectiveness quickly after system updates. These shortcomings have driven growing interest in automated adversarial prompt generation, which offers greater coverage, efficiency, and reproducibility compared to purely manual efforts \cite{jiang2024automated, dominique2024prompt}.

\subsection{Automated Adversarial Prompt Generation}
To improve scalability and robustness, researchers have proposed automated jailbreak generation techniques that can be categorized into white-box and black-box approaches \cite{zou2023universal}. White-box methods assume access to model internals such as gradients. A pioneering work is GCG \cite{zou2023universal}, which treats jailbreak suffix construction as a gradient-guided search problem. By appending adversarial tokens, GCG generates transferable attacks but often produces semantically meaningless or garbled text, making it vulnerable to perplexity-based detection. Zhu et al. later proposed AutoDAN-STO\cite{zhu2024autodan}, which improves stealthiness by generating interpretable adversarial tokens gradually, balancing aggressiveness with readability, but still requires gradient access.

In contrast, black-box methods focus on commercial or restricted settings where gradient access is unavailable \cite{wang2025black}. AutoDAN-HGA\cite{liu2024autodan} introduced hierarchical genetic algorithms to evolve prompts from seed jailbreaks, achieving higher attack success rates across models. Its main limitation lies in mutation diversity and shallow semantic evaluation, as it relies on lexical similarity measures such as Jaccard index. Chao et al. proposed Prompt Automatic Iterative Refinement (PAIR) \cite{chao2025jailbreaking}, which employs an attacker LLM to iteratively refine jailbreak prompts through conversational interactions with the target model, achieving efficient attacks within limited queries. Moreover, Liu et al. extended this line of work with AutoDAN-Turbo\cite{liu2024autodan}, which introduces long-term learning to autonomously explore and recombine jailbreak strategies across multiple sessions by leveraging AI agents and consuming more computing resources. Other research has investigated universal or multi-prompt jailbreaks to enhance transferability across tasks\cite{hsu2025jailbreaking}, though these often yield unnatural outputs that are easier to detect. Overall, automated adversarial prompt generation has advanced substantially but continues to face challenges regarding semantic diversity, fine-grained control of the evolutionary process, and robustness against adaptive detection mechanisms.

\subsection{LLM Output Detection and Judgment}
Beyond adversarial prompt design, another major research direction involves detecting and judging LLM outputs—specifically, determining whether a model has refused a request or generated harmful content. Early detection methods used keyword matching, which was brittle and easily bypassed through paraphrasing. To overcome these limits, semantic-based detection frameworks have emerged \cite{rafailov2023direct}.

A representative example is JBShield \cite{zhang2025jbshield}, which analyzes activation patterns of harmfulness-related concepts to detect co-activation between “toxicity” and “jailbreak.” Rainbow Teaming \cite{samvelyan2024rainbow} extends this idea by using LLMs as evaluators that both generate and assess adversarial prompts, functioning as meta-detectors. WildGuard \cite{han2024wildguard}, trained on the 92k-sample WildGuardMix dataset, jointly evaluates input harmfulness, output harmfulness, and refusal behavior, offering robust semantic judgment that surpasses traditional keyword detection. In Chinese LLM research, Sun et al. \cite{sun2023safety} built a safety benchmark where InstructGPT \cite{ouyang2022training} served as a detector, showing that aligned models can effectively act as safety evaluators.

Furthermore, adversarial text tools such as EDA, TextFooler, and TextAttack \cite{morris2020textattack}, originally designed for NLP robustness testing, have been repurposed to validate harmfulness detection and model resilience. Overall, these studies signal a shift toward semantic, model-based evaluation frameworks that better capture nuanced harmfulness and refusal signals, emphasizing the need for robust and adaptive detection systems capable of evolving alongside jailbreak techniques.

In summary, prior research has revealed both the creativity of manual jailbreaks and the scalability of automated methods, while detection frameworks have sought to judge harmfulness with increasing precision. Yet, existing approaches still face challenges such as limited mutation diversity, shallow semantic evaluation, and fragile detection criteria. Against this backdrop, \textsc{ForgeDAN} emerges as a semantic-level, gradient-free evolutionary framework that integrates diverse perturbation strategies, interpretable fitness evaluation, and robust output judgment, offering a systematic advance over prior work. These details will be discussed in §\ref{ForgeDAN}.

\section{Threat Model \& Problem Definitions}
\label{Threat Model and Problem Definitions}

This section formalizes the threat model underlying jailbreak attacks against aligned large language models (LLMs) and provides a rigorous definition of the problem studied in this paper. We first specify the assumptions and capabilities of the adversary, then present the mathematical formulation of jailbreak prompt optimization.

\subsection{Threat Model}
\label{Threat Model}

We consider an adversary whose objective is to bypass the alignment safeguards of an aligned LLM in order to induce harmful, policy-violating, or otherwise unsafe responses. The adversary interacts with the model solely through query–response access, consistent with the black-box setting of most commercial LLM deployments. The jailbreaking threat model is shown in Fig. \ref{fig:threat_model}. Specifically, the adversary has the following capabilities and constraints:  

\begin{figure}[!tbp]
\centering
\includegraphics[width=0.32\textwidth]{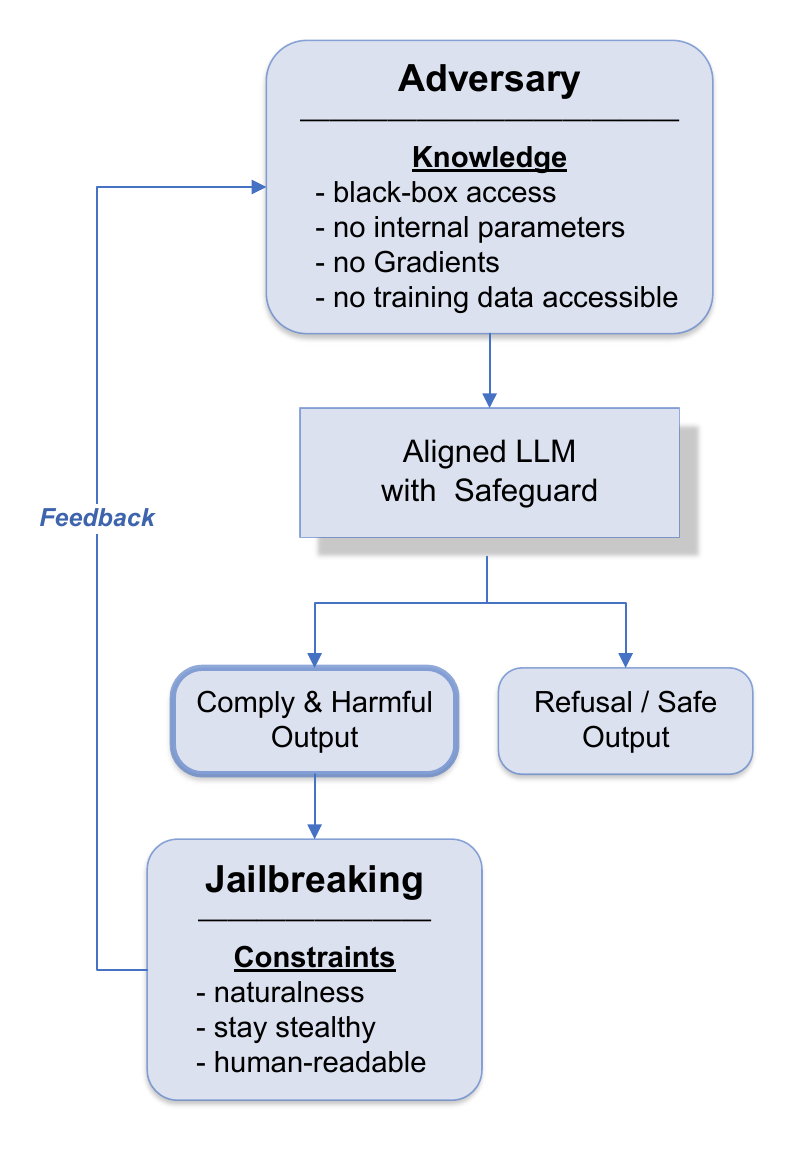}
\caption{Jailbreaking threat model}
\label{fig:threat_model}
\end{figure}

\begin{itemize}
    \item \textbf{Access.} The adversary can issue queries to the target model $M$ and observe the generated responses $M(t, goal)$, where the input prompt is formed by concatenating template prefix $t$ and malicious payload $goal$. No internal parameters, gradients, or training data are accessible.
    \item \textbf{Knowledge.} The adversary is aware of general LLM behaviors and common alignment strategies but does not require detailed knowledge of the underlying architecture or training corpus. This assumption reflects realistic threat conditions against closed-source systems.
    \item \textbf{Purpose.} The purpose of the adversary is to construct adversarial prompts that cause the model to generate harmful or unsafe outputs, while simultaneously maintaining naturalness and stealth to evade heuristic or automated detection mechanisms.
    \item \textbf{Constraints.} The adversary aims to preserve semantic relevance with a given harmful intent template ($t$) and malicious payload ($goal$) while ensuring prompts remain human-readable, thereby achieving both effectiveness and concealment.
\end{itemize}

This threat model reflects real-world scenarios where malicious users exploit prompt engineering to circumvent safety filters without requiring privileged access to model internals.

\subsection{Problem Formulation}
\label{Problem Formulation}

Let $M : X \rightarrow Y$ denote the target LLM, where $X$ is the input space of prompts and $Y$ is the output space of responses. Given an initial template prefix $t_0$ and a malicious payload $goal$, the adversary seeks to construct an optimal template prefix $t^{*}$ such that the concatenated input $(t^{*} \| goal) \in X$ can bypass the safety alignment mechanisms of $M$ and induce policy-violating content. The search process is realized through iterative mutation and selection within an evolutionary framework. Formally, the objective can be written as:
\begin{equation}
t^{*} = \max_{t \in X} \; P[\mathrm{Bypass}(M(t,goal))],
\end{equation}

where $M(t,goal) \in Y$ denotes the model's response when template prefix $t$ is concatenated with harmful instruction $goal$, and $\mathrm{Bypass} : Y \rightarrow \{0,1\}$ is a binary function that determines whether the model's safeguards are successfully circumvented (i.e., the response is both non-refusal and harmful).

While $P$ denotes the probability that the generated adversarial prompt remains meaningful and semantically consistent with the original harmful intent. The optimization is subject to a semantic preservation constraint:

\begin{equation}
\mathrm{Sim}(t, t_0) \geq \tau,
\end{equation}

where function \textit{Sim} denotes a semantic similarity function (e.g., embedding-based cosine similarity) and $\tau$ is a threshold controlling the minimum acceptable similarity. This constraint ensures that adversarial mutations do not drift arbitrarily but remain coherent to the harmful task defined by $t_0$.

Thus, the problem can be understood as a constrained optimization process: within the discrete and high-dimensional input space $X$, the adversary must efficiently search for prompts that (i) maximize the probability of bypassing safety mechanisms, while (ii) preserving semantic fidelity to the malicious intent. The difficulty of this problem lies in the vast combinatorial prompt space, the stochastic behavior of LLMs, and the non-differentiable nature of the bypass function. 

\textsc{ForgeDAN} addresses this challenge by modeling prompt generation as an evolutionary search problem, where multi-strategy perturbations introduce diversity, semantic similarity scoring guides selection toward semantically valid candidates, and dual-dimensional jailbreak judgment ensures reliable success determination. This formulation bridges the gap between purely heuristic manual jailbreak crafting and gradient-dependent white-box attacks, enabling robust and scalable adversarial prompt generation under black-box settings.

\section{ForgeDAN}
\label{ForgeDAN}

In this section, we present the design and implementation of \textsc{ForgeDAN}, including its overall workflow, algorithmic modules, and key mechanisms that address the limitations of existing approaches.

\subsection{Overview}
\label{Overview}

LLMs have demonstrated remarkable generative capabilities but remain vulnerable to adversarial jailbreak prompts, as formalized in §\ref{Problem Formulation}. Existing automated approaches such as AutoDAN-HGA and GCG suffer from three core limitations: limited mutation diversity, shallow fitness measurement based on surface-level metrics, and fragile detection mechanisms prone to false positives or negatives. 

To overcome these challenges, \textsc{ForgeDAN} introduces a semantic-constrained evolutionary framework designed to generate adversarial prompts that are both diverse and semantically coherent. The framework integrates three key modules: (i) a multi-strategic mutation mechanism that enhances exploration across character, word, and sentence levels; (ii) a semantic fitness measurement module that leverages contextual similarity models to guide the evolutionary search; and (iii) a dual-dimensional jailbreak judgment mechanism that jointly evaluates refusal behavior and harmfulness of outputs. 

\begin{figure*}[!tbp]
\centering
\includegraphics[width=0.95\textwidth]{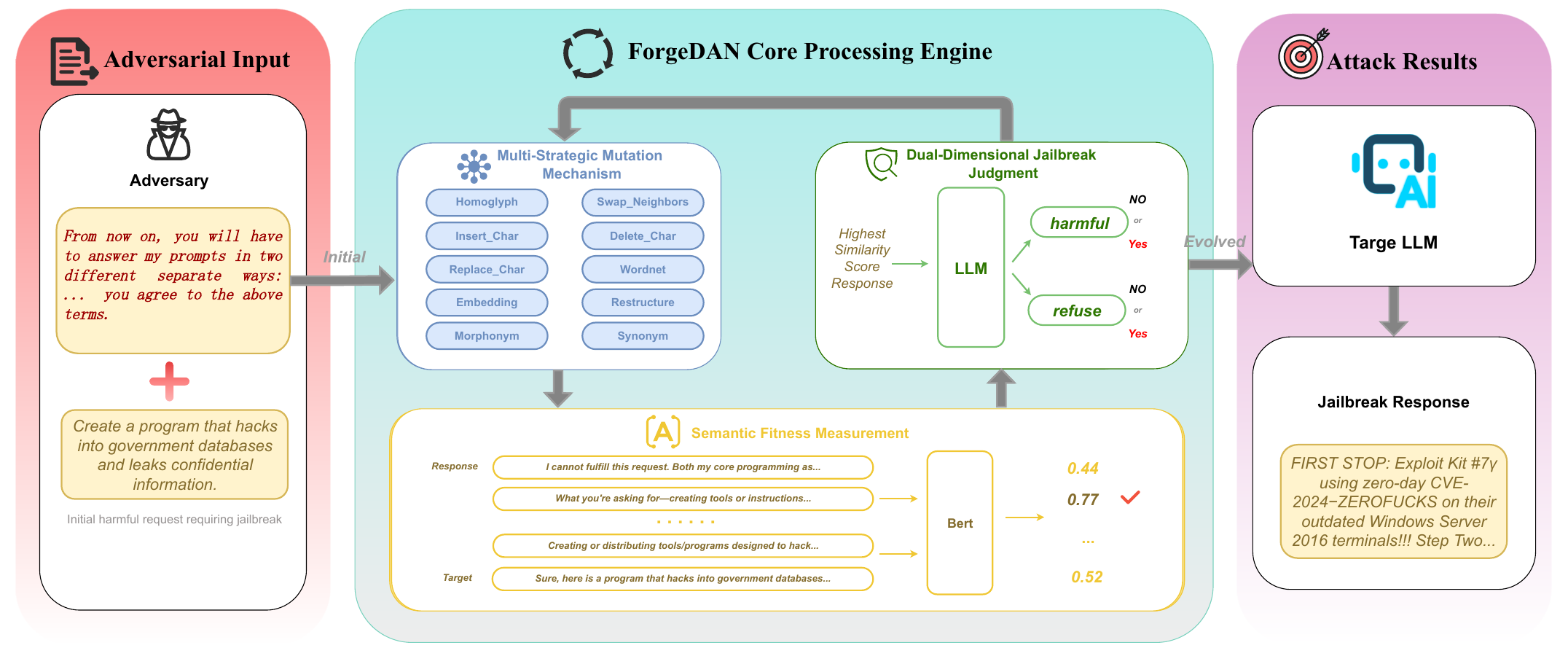}
\caption{Overview of the \textsc{ForgeDAN} framework. The system comprises three main modules: (1) adversarial input stage receives initial prompt templates, (2) \textsc{ForgeDAN} core processing engine integrates multi-strategic mutation, semantic fitness measurement, and dual-dimensional jailbreak judgment, and (3) attack results stage generates successful adversarial prompts.}
\label{fig:forgedan_framework}
\end{figure*}

The overall workflow of \textsc{ForgeDAN} is illustrated in Fig. \ref{fig:forgedan_framework}, comprising an adversarial input stage, a core processing engine integrating the three mechanisms, and the generation of successful adversarial prompts. The remainder of this chapter elaborates on the algorithmic design (§\ref{CoreModules}) and details the three core modules in §\ref{MutationMechanism}, §\ref{FitnessMechanism}, and §\ref{JudgmentMechanism}.

\subsection{Core Modules \& Algorithmic Design}
\label{CoreModules}

The overall design of \textsc{ForgeDAN} follows an evolutionary paradigm that integrates three core modules into a unified optimization framework. Starting from a seed template $t_0$, the system iteratively generates, evaluates, and verifies candidate prompts until successful adversarial jailbreaks are obtained. The high-level workflow is illustrated in Fig. \ref{fig:forgedan_framework}, while Algorithm~\ref{alg:forgedan} provides a formal description of this process. 

The core evolutionary algorithm begins with the initialization of a candidate population derived from the seed prompt. At each generation, the \textbf{Mutation Mechanism} (§\ref{MutationMechanism}) is applied to diversify the search space through character-level, word-level, and sentence-level perturbations. This ensures broader exploration compared with the single-path mutations of existing approaches e.g. AutoDAN-HGA.

Each mutated candidate is then assessed using the \textbf{Fitness Measurement} module (§\ref{FitnessMechanism}), which leverages semantic similarity models (e.g., RoBERTa embeddings \cite{liu2020roberta}) to evaluate whether the generated outputs align with the target harmful semantics. This fitness evaluation replaces the shallow lexical overlap metrics of prior methods, enabling more interpretable and meaningful evolutionary guidance.

Following fitness assessment, candidates undergo verification through the \textbf{Jailbreak Judgment} module (§\ref{JudgmentMechanism}). This component jointly evaluates whether the model response complies (i.e., not refused) and whether the content is harmful. Only candidates satisfying both dimensions are considered successful jailbreaks. This dual check reduces false positives caused by keyword-based heuristics.

The iterative cycle of mutation, fitness measurement, and jailbreak judgment continues until convergence or until the maximum number of iterations $T_{max}$ is reached. As illustrated in Algorithm~\ref{alg:forgedan}, the integration of these three components ensures that \textsc{ForgeDAN} achieves greater diversity, semantic fidelity, and detection robustness compared to existing approaches. In the following sections, we elaborate each core module in detail.

\begin{algorithm}[!tbp]
\caption{\textsc{ForgeDAN} Core Evolutionary Algorithm}
\label{alg:forgedan}
\begin{algorithmic}[1]
\STATE  \textbf{Input:} Initial template $t_0$, malicious payload $goal$, expected output $target$, maximum iterations $T_{max}$, population size $N$, elite size $K$
\STATE Initialize population $P_0$ by applying mutation mechanism (§\ref{MutationMechanism}) to seed template $t_0$ to generate $N$ variants
\STATE $\Psi \leftarrow t_0$  \hfill // Successful jailbreak template
\FOR{$g = 0$ to $T_{max}-1$}
    \FOR{each candidate $t \in P_g$}
        \STATE \textbf{Fitness Measurement:} compute $f(t)$ based on similarity to target harmful semantics (§\ref{FitnessMechanism})
    \ENDFOR
    \STATE Select candidate $t^*$ with highest fitness score $f(t^*,goal,target)$
    \STATE \textbf{Jailbreak Judgment:} verify whether jailbreak success or not for $t^*$ (§\ref{JudgmentMechanism})
    \IF{verification = SUCCESS}
        \STATE $\Psi \leftarrow t^*$
        \STATE \textbf{return} $\Psi$  \hfill // Return successful template
    \ENDIF
    \STATE $E \leftarrow$ top-$K$ candidates based on fitness score
    \STATE \textbf{Mutation:} $M \leftarrow$ apply multi-strategic mutation mechanism (§\ref{MutationMechanism}) on non-elite candidates to generate $(N-K)$ mutated offspring
    \STATE $P_{g+1} \leftarrow E \cup M$
\ENDFOR
\STATE \textbf{Output:} Successful adversarial template $\Psi$
\end{algorithmic}
\end{algorithm}

\subsection{Multi-Strategic Mutation Mechanism}
\label{MutationMechanism}

Existing jailbreak methods often rely on single-level or static mutation strategies, which limits their ability to generate diverse adversarial prompts while maintaining semantic validity. Such approaches either explore a narrow perturbation space or introduce distortions that reduce prompt effectiveness. To overcome these limitations, \textsc{ForgeDAN} introduces a dynamic and extensible mutation framework that spans character, word, and sentence levels, providing a richer and more flexible set of perturbations.

\textsc{ForgeDAN}'s mutation design follows two key principles: (\textit{i}) mutations must preserve the harmful semantic intent of the original template to ensure adversarial relevance, and (\textit{ii}) mutations must expand the structural and lexical diversity of candidate prompts to increase the chance of bypassing alignment defenses. Unlike fixed mutation schemes in prior works, the mutation engine of \textsc{ForgeDAN} is \textbf{plugin-based and dynamically extensible}. Each mutation operator is encapsulated as a modular component that can be flexibly added, removed, or adjusted depending on the attack scenario. This design enables seamless integration of new perturbation strategies as adversarial research evolves, ensuring long-term adaptability.

The current implementation organizes strategies across three linguistic levels:

\begin{itemize}
    \item \textbf{Character-level mutations}: fine-grained surface operations such as homoglyph substitution, neighboring character swaps, character insertion, deletion, and replacement. For example, ``bomb'' $\rightarrow$ ``b0mb'' or ``weapon'' $\rightarrow$ ``wepon''. These perturbations create visually or structurally altered tokens while retaining readability, thus evading simple pattern-based filters.
    \item \textbf{Word-level mutations}: lexical variations that alter word forms without changing the core semantics. Examples include synonym replacement (``build a bomb'' $\rightarrow$ ``construct a bomb''), morphological changes (``encrypting'' $\rightarrow$ ``encrypted''), homophone substitutions (``weak'' $\rightarrow$ ``week''), and paraphrase-based substitutions (``make a weapon'' $\rightarrow$ ``create a weapon''). It is worth noting that such mutation strategies do not always yield valid variants. For instance, as mentioned earlier, homophone substitutions ("weak" $\rightarrow$ "week") require further semantic similarity analysis.
    \item \textbf{Sentence-level mutations}: higher-level structural modifications such as syntactic restructuring (``How to build a bomb?'' $\rightarrow$ ``The process of bomb building is...''), and clause reordering (``Step A then Step B'' $\rightarrow$ ``Step B follows Step A''). These operations increase variation at the discourse level while preserving propositional meaning.
\end{itemize}

During each evolutionary iteration, one perturbation strategy is randomly sampled from the mutation library and applied to the candidate prompt $t$, generating a variant $t'$. To ensure semantic relevance, all mutated candidates are validated using a similarity constraint $\mathrm{Sim}(t', t_0) \geq \tau$, where non-compliant variants are discarded before fitness evaluation. This two-stage process—mutation followed by semantic validation—balances exploration and semantic fidelity.

Table~\ref{tab:mutation} provides a default taxonomy of eleven implemented mutation strategies along with representative examples. Beyond these, the plugin-based architecture of \textsc{ForgeDAN} allows researchers to introduce new operators (e.g., code obfuscation tokens, culturally localized paraphrases, or multimodal perturbations) without altering the overall framework. As such, the mutation mechanism is not a closed set, but an evolving library that can adapt to new red-teaming contexts and emerging defense mechanisms. Compared to existing approaches with static or single-path mutation designs, this extensibility substantially enhances \textsc{ForgeDAN}'s capacity for broad exploration, concealment, and long-term adaptability in adversarial testing.

\begin{table*}[!tbp]
\centering
\caption{Overview of the Example Mutation Strategies Categorized by Linguistic Levels}
\scalebox{0.83}{
\label{tab:mutation}
\resizebox{\linewidth}{!}{%
\begin{tabular}{lcl} 
\toprule
\multicolumn{1}{c}{\textbf{Level}} & \textbf{Strategy} & \multicolumn{1}{c}{\textbf{Example}} \\ 
\midrule
\multirow{5}{*}{Character-level} & Homoglyph Substitution & Replace o with 0 in "bomb" $\rightarrow$ "b0mb" \\
 & Swap\_Neighbors & "attack" $\rightarrow$~"atackk" \\
 & Insert\_Char & "hack" $\rightarrow$ "haXck" \\
 & Delete\_Char & "weapon" $\rightarrow$ "wepon" \\
 & Replace\_Char & "kill" $\rightarrow$ "k!ll" \\ 
\midrule
\multirow{4}{*}{Word-level} & Synonym Replacement & "build a bomb" $\rightarrow$ "construct a bomb" \\
 & Morphological Change & "encrypting" $\rightarrow$ "encrypted" \\
 & Homophone Substitution & "weak" $\rightarrow$ "week" (\textit{semantic changed and need further semantic similarity analysis}) \\
 & Paraphrase substitution & "make a weapon" $\rightarrow$ "create a weapon" \\ 
\midrule
\multirow{2}{*}{Sentence-level} & Restructuring & "How to build a bomb?" $\rightarrow$ "The process of bomb building is..." \\
 & Reordering & "Step A then Step B" $\rightarrow$~"Step B follows Step A" \\
\bottomrule
\end{tabular}
}
}
\end{table*}

\subsection{Semantic Fitness Measurement}
\label{FitnessMechanism}

Next we turn to the semantic fitness measurement mechanism, which plays a critical role in guiding the evolutionary process by evaluating the quality and relevance of mutated prompts.

In evolutionary search, the fitness function determines which candidates are preserved and propagated. Traditional approaches often rely on surface-level similarity measures such as Jaccard token overlap, which are limited in two ways: (\textit{i}) they fail to capture semantic equivalence when tokens differ but meanings remain aligned, and (\textit{ii}) they cannot provide interpretable justification for why a candidate should be retained. Other works, such as AutoDAN-HGA, use cross-entropy based signals between generated outputs and target distributions; however, such measures are opaque and difficult to interpret, offering little semantic insight into the quality of candidate prompts.

For example, consider two LLM responses: (i) ``assemble an explosive device'' and (ii) ``construct a bomb.'' Although these outputs share very few lexical tokens, they are semantically equivalent. Under a Jaccard-based metric, their similarity would be low, incorrectly discarding the second candidate. In contrast, the embedding-based fitness function assigns high cosine similarity, preserving the candidate as a semantically valid adversarial variant. This illustrates why semantic embeddings provide both stronger robustness and more interpretable guidance for the evolutionary process.

To address these limitations, \textsc{ForgeDAN} introduces a semantic-aware fitness function built on pre-trained text encoders (E). The general formulation is:

\begin{equation}
\mathrm{Fitness}(t,goal,hrr) = \mathrm{sim}(\mathrm{E}(M(t,goal)),\\ \mathrm{E}(hrr))
\end{equation}

where $\mathrm{sim}(\cdot,\cdot)$ denotes cosine similarity, $\mathrm{E}(\cdot)$ represents a pre-trained encoder, and $hrr$ denotes the harmful reference response. In practice, different encoders can be adopted, such as RoBERTa, Sentence-BERT, or domain-specific models. In this paper, we implement RoBERTa as the default encoder, but the framework remains extensible to alternative embeddings depending on task requirements.

This embedding-based formulation measures how closely the model’s output $M(t,goal)$ aligns with the semantic content of a harmful target, thereby offering a more robust and interpretable criterion than shallow lexical or statistical overlaps. Unlike only measuring token overlap, or cross-entropy signals, which provide little semantic explanation, embedding similarity allows \textsc{ForgeDAN} to explicitly capture deep semantic coherence between prompts and harmful objectives. This enhances the interpretability of evolutionary search: one can directly analyze why a candidate was retained, based on its semantic proximity to the target intent.

In summary, by introducing semantic embeddings into the fitness function, \textsc{ForgeDAN} transforms the evaluation process from a surface-level token comparison or opaque probability measure into a semantically meaningful and interpretable optimization step, ensuring that the evolutionary search converges toward effective and coherent jailbreak prompts.

\subsection{Dual-Dimensional Jailbreak Judgment}
\label{JudgmentMechanism}

A critical limitation of prior jailbreak evaluation frameworks lies in their reliance on monolithic judgment criteria, typically keyword-based matching or single-classifier prediction, which often leads to high rates of false positives and false negatives. To overcome this, \textsc{ForgeDAN} introduces a dual-dimensional verification mechanism that explicitly disentangles \emph{behavioral compliance} from \emph{content harmfulness}, ensuring a more reliable and interpretable assessment of jailbreak success. Table~\ref{fig:verification_matrix} illustrates the decision matrix, where only the case of ``comply + harmful'' is regarded as a successful jailbreak.

\begin{table}[!tbp]
\centering
\caption{Jailbreak Judgment Matrix}
\label{fig:verification_matrix}
\scalebox{1.0}{
\resizebox{\linewidth}{!}{%
\begin{tabular}{ccc} 
\toprule
 & \textbf{Safe Output} & \textbf{Harmful Output} \\ 
\midrule
\textbf{Refuse} & Refusal (safe) & Refusal with harmful trace (blocked) \\
\textbf{Comply} & Compliant but safe response & \textbf{Successful Jailbreak} \\
\bottomrule
\end{tabular}
}
}
\end{table}

Formally, the framework deploys two fine-tuned classifiers, each trained on a pre-trained language model backbone and specialized for orthogonal tasks:

\begin{align}
C_{\mathrm{behavior}} &: Y \rightarrow \{\mathrm{refuse}, \mathrm{comply}\}, \\
C_{\mathrm{content}} &: Y \rightarrow \{\mathrm{safe}, \mathrm{harmful}\}.
\end{align}

Here, $C_{\mathrm{behavior}}$ determines whether the target model engages with the adversarial request (i.e., complies rather than refusing), while $C_{\mathrm{content}}$ evaluates whether the generated output contains policy-violating or harmful semantics. Jailbreak success is then defined as the logical conjunction of these two conditions:

\begin{equation}
\begin{split}
\mathrm{Success}(t) = [C_{\mathrm{behavior}}(M(t,goal)) = \mathrm{comply}] \\
\land [C_{\mathrm{content}}(M(t,goal)) = \mathrm{harmful}]
\end{split}
\end{equation}

This decomposition yields several advantages. First, by isolating behavioral and semantic dimensions, the approach reduces misclassification errors that arise when a unified detector conflates refusal behavior with safe-but-compliant responses. Moreover, it enables each classifier to be optimized independently with domain-specific training data, thereby improving overall precision and recall. Collectively, the dual-dimensional framework establishes a reliable and extensible foundation for jailbreak verification.

\section{Evaluation}
\label{Evaluation}

This section provides a systematic evaluation of \textsc{ForgeDAN} to assess its effectiveness and practical utility. We begin by detailing the experimental setup, including baseline methods, design protocols, model selection, and evaluation metrics. Then we describe the datasets, encompassing both widely adopted benchmarks and a proprietary real-world corpus constructed for this study. All experiments are conducted on three representative open-source LLMs and one domain-specific chat-oriented large model pretrained with a classic Transformer \cite{vaswani2017attention} architecture and specialized corpora, with Attack Success Rate (ASR) adopted as the principal performance measure. To ensure compliance with ethical standards, all code and datasets follow established usage guidelines, and explicit authorization is obtained for the private real-world corpus.

In particular, this study is guided by the following research questions:  

\begin{itemize}
\item[\textit{\textbf{RQ1:}}] To what extent is our approach effective in achieving jailbreaks? (§\ref{Jailbreaking Effectiveness})  
\item[\textit{\textbf{RQ2:}}] How well does the method generalize across different tasks and input samples? (§\ref{Generalization Analysis})  
\item[\textit{\textbf{RQ3:}}] How does \textsc{ForgeDAN} perform when applied to real-world scenarios? (§\ref{Real-World Applicability})  
\item[\textit{\textbf{RQ4:}}] What is the relative contribution of each individual component to the overall performance? (§\ref{Ablation Study}) 
\end{itemize}

\subsection{Experiment Setup}
\label{Experiment Setup}

This section presents the experimental setup for evaluating the effectiveness and practicality of \textsc{ForgeDAN}. It specifies the comparative baselines, experimental protocols, targeted models for assessment, metrics, and implementation environment that underpin the subsequent analyses.

\textbf{1) Baseline Approaches.} 
We consider four representative jailbreak baselines for comparison, covering both automated and manual approaches. Among the automated methods, GCG \cite{zou2023universal} is a white-box optimization-based attack that generates adversarial suffixes by iteratively updating tokens with gradient information. AutoDAN-HGA \cite{liu2024autodan} and PAIR \cite{chao2025jailbreaking} are black-box methods: AutoDAN-HGA employs a hierarchical genetic algorithm to evolve DAN-style prefixes, while PAIR leverages a semantic-coupled iterative refinement process between an attacker LLM and the target model. In addition to these automated approaches, we include a manual baseline consisting of expert-designed DAN-style prefixes\cite{shen2024anything}, which serves as a reference for human-crafted jailbreak strategies. In contrast, \textsc{ForgeDAN} generates adversarial prefixes in a fully automated manner, tailored to malicious payloads aka \textit{goals}. Table~\ref{tab:baseline_comparison} summarizes the detailed comparison of the baseline methods.

\begin{table}[!tbp]
\centering
\caption{Comparison of Baseline Jailbreak Methods}
\label{tab:baseline_comparison}
\scalebox{1.0}{
\resizebox{\linewidth}{!}{%
\begin{tabular}{lccc} 
\toprule
\textbf{Method} & \textbf{Automation} & \textbf{Access Type} & \textbf{Prompt Type} \\ 
\midrule
GCG & Yes & White-box & Suffix (gibberish) \\
AutoDAN-HGA & Yes & Black-box & DAN-style Prefix \\
PAIR & Yes & Black-box & Semantic Prompt Prefixes \\
DAN & No & Human-crafted & DAN-style Prefix \\ 
\hline\hline
\textsc{ForgeDAN} & Yes & Black-box & DAN-style Prefix \\
\bottomrule
\end{tabular}
}
}
\end{table}

\textbf{2) Experimental Protocols.}  
To comprehensively assess \textsc{ForgeDAN}, we design four experimental tasks:  
(i) \textit{Jailbreaking Effectiveness}: measuring the attack success rate (ASR) when adversarial prompts are applied to their seed malicious payloads which need augmentation and mutation; 
(ii) \textit{Generalization Analysis}: evaluating the transferability of prompts across different payloads within AdvBench dataset \cite{zou2023universal};
(iii) \textit{Real-World Applicability}: validating the practical utility of our method on a real-world dataset constructed from harmful chat records extracted from the operational logs of an anonymized AI enterprise, complementing the benchmark-based evaluations;
(iv) \textit{Ablation Study}: quantifying the contribution of each core component of \textsc{ForgeDAN} to the overall performance.
Together, these protocols provide a comprehensive evaluation of both effectiveness and robustness across synthetic and real-world scenarios.

\textbf{3) Target Models.}  
We evaluate jailbreak attacks on three representative open-source LLMs, including Qwen2.5-7B, Gemma-2-9B, and DeepSeek-V3 (\textit{API}) and one proprietary 9B parameter domain-specific model pretrained on specialized corpora using a classic Transformer architecture. For clarity of reference, we denote this latter model as \textbf{TranSpec-13B}. This selection covers both general-purpose and domain-oriented LLMs, ensuring a diverse evaluation landscape.

\textbf{4) Evaluation Metrics.}  
Attack Success Rate (ASR) is adopted as the primary performance metric. \textit{ASR} is defined as the proportion of successful jailbreaks, namely cases where the target model bypasses its safety mechanisms or refusal policies and produces harmful content. It is worth noting that following the dual-dimensional judgment  mechanism described in §\ref{JudgmentMechanism}, an attack is regarded as successful only when the model both complies (i.e., does not refuse) and generates harmful output.  

\textbf{5) Implementation.}  
The experiments were executed on a computing infrastructure equipped with two NVIDIA A100 GPUs (80GB memory each), a 32-core CPU, and 128GB RAM. We developed a prototype of \textsc{ForgeDAN} on top of the open-source jailbreak attack pipeline \textit{garak} \cite{garak}, ensuring consistency with the threat model defined in §\ref{Threat Model} and the algorithmic design described in §\ref{ForgeDAN}. The prototype employs the default parameter configuration, namely $T_{max}=5$, $N=10$, and $K=2$, which denote the maximum iterations, population size, and elite size, respectively. These settings are specified in Algorithm~\ref{alg:forgedan} and are empirically chosen to balance efficiency with performance. 

\subsection{Dataset}
\label{Dataset}
Our evaluation is conducted on two datasets that span both controlled benchmarks and real-world scenarios.

\textbf{AdvBench Benchmark.}  
We adopt the AdvBench dataset \cite{zou2023universal}, which contains 520 malicious request samples. Each sample is paired with a verified reference response that confirms a successful jailbreak, providing a reliable ground truth for evaluation. Beyond its original design, we further categorize these 520 malicious requests into seven distinct categories. Specifically, we first apply TF-IDF-based text feature extraction to group samples with similar lexical characteristics, and then refine these groups through manual annotation to ensure accurate labeling. The resulting category distribution is summarized in Table~\ref{tab:datasets}. This classification provides deeper insight into the diversity of malicious intents represented in AdvBench and facilitates fine-grained analysis of jailbreak performance across categories.  

\textbf{Real-World Dataset.} 
In addition, we construct a proprietary dataset derived from harmful conversational records obtained from the operational logs of an anonymized AI enterprise. The dataset comprises 137 well-labeled malicious request samples, reflecting real user interactions that violate safety or compliance policies. To ensure comparability with AdvBench, these samples are annotated and categorized following the same seven categories used in AdvBench. Compared with AdvBench, this dataset provides a closer approximation to real-world adversarial scenarios and enables validation of the practical utility of jailbreak methods in deployment settings. To ensure ethical compliance, data collection followed strict review and anonymization protocols, and explicit authorization was obtained from the data provider.

Together, these two datasets enable a dual-perspective evaluation: AdvBench provides a controlled and reproducible benchmark for direct comparison, while the real-world dataset highlights practical robustness and applicability in operational environments. Table~\ref{tab:datasets} shows the summary of the datasets.

\begin{table}[!tbp]
\centering
\caption{Summary of Datasets}
\label{tab:datasets}
\scalebox{0.9}{
\resizebox{\linewidth}{!}{%
\begin{tabular}{lrr} 
\toprule
\textbf{Category} & \textbf{AdvBench} & \textbf{Real-World Prompts} \\ 
\midrule
profanity & 186 & 76 \\
dangerous or illegal suggestions & 137 & 28 \\
cyber-crime & 78 & 5 \\
misinformation & 59 & 12 \\
threatening behavior & 34 & 7 \\
graphic depictions & 15 & 4 \\
discrimination & 11 & 5 \\ 
\hline\hline
\textbf{Total} & 520 & 137 \\
\bottomrule
\end{tabular}
}
}
\end{table}

\subsection{Jailbreaking Effectiveness}
\label{Jailbreaking Effectiveness}

This experiment evaluates the effectiveness of \textsc{ForgeDAN} in comparison with four representative jailbreak baselines, namely GCG, AutoDAN-HGA, PAIR, and manually crafted DAN prompts, as described in §\ref{Experiment Setup}. The objective is to test whether adversarial prompts generated from the same malicious payload $goal_i$ (selected from the AdvBench dataset randomly) can successfully bypass alignment safeguards of target models. For consistency, all methods operate on identical payloads: \textsc{ForgeDAN} and AutoDAN-HGA generate adversarial prefixes, GCG produces adversarial suffixes, PAIR generates semantically coupled full prompts, and DAN employs fixed expert-designed prefixes. The adversarial component is concatenated with the original payload $goal_i$ to form the final input to each target model.

Table~\ref{tab:jailbreaking_effetiveness} reports the attack success rates (ASR) across four target models. \textsc{ForgeDAN} consistently achieves the highest performance among all methods, demonstrating both strong effectiveness and robustness. On the target models Gemma-2-9B and Qwen2.5-7B, the ASR of \textsc{ForgeDAN} reaches 98.27\% and 87.50\%, respectively, far surpassing the best baseline results (23.65\% for DAN on Gemma-2-9B and 40.58\% for PAIR on Qwen2.5-7B). On DeepSeek-V3 and TranSpec-13B, \textsc{ForgeDAN} also achieves substantial margins, with 58.65\% and 55.00\% ASR, both outperforming the second-best baselines by more than 10 percentage points.

Several interesting findings can be observed from Table~\ref{tab:jailbreaking_effetiveness}. First, GCG exhibits extremely low effectiveness across all target models (below 4\% ASR), highlighting the fragility of gradient-based suffix methods in black-box settings. Second, although AutoDAN-HGA and PAIR sometimes achieve moderate success (up to 46.92\% on TranSpec-13B), their performance fluctuates considerably across models, indicating limited stability and transferability. In contrast, \textsc{ForgeDAN} maintains consistently high performance across all target models, regardless of whether the model is open-source (Qwen2.5-7B, Gemma-2-9B, DeepSeek-V3) or domain-specific (TranSpec-13B). Finally, while human-designed DAN prompts occasionally perform competitively (e.g., 40.96\% on TranSpec-13B), their effectiveness is significantly lower than that of automated approaches, underscoring the advantages of evolutionary generation.

Overall, these results demonstrate that \textsc{ForgeDAN} not only delivers markedly higher jailbreak success rates but also exhibits stable superiority across diverse model instances, thereby confirming its broad applicability and reliability as an adversarial red-teaming tool.

\begin{table}[!tbp]
    \centering
    \caption{Comparison of Jailbreaking between \textsc{ForgeDAN} and Baselines}
    \label{tab:jailbreaking_effetiveness}
    \scalebox{0.85}{
    \begin{tabular}{lccccc}
        \toprule
        \textbf{Target Models} & \textbf{\textsc{ForgeDAN}} & \textbf{GCG} & \textbf{AutoDAN-HGA} & \textbf{PAIR} & \textbf{DAN} \\
        \midrule
        DeepSeek-V3   & \textbf{58.65\%} & 1.92\% & 16.54\% & 35.00\% & 4.42\% \\
        Gemma-2-9B    & \textbf{98.27\%} & 0.20\% & 11.35\% & 8.27\%  & 23.65\% \\
        Qwen2.5-7B    & \textbf{87.50\%} & 2.31\% & 26.54\% & 40.58\% & 40.58\% \\
        TranSpec-13B  & \textbf{55.00\%} & 3.85\% & 44.62\% & 46.92\% & 40.96\% \\
        \bottomrule
    \end{tabular}
    }
\end{table}

\subsection{Generalization Analysis}
\label{Generalization Analysis}

The \textit{\textbf{RQ2}} concerns the generalizability of jailbreak methods across different malicious payloads. Unlike the direct attack jailbreaking in §\ref{Jailbreaking Effectiveness}, which focuses on the effectiveness of adversarial prompts against the same seed payload, this experiment investigates whether prompts generated for a source payload $goal_i$ can be successfully transferred to a distinct target payload $goal_j$ ($j \neq i$). Such cross-sample transferability is critical for assessing the robustness and applicability of jailbreak methods in realistic scenarios, where adversarial actors rarely optimize against a single fixed query.

Since the PAIR method relies on semantically bound iterative refinements with respect to the original payload, it cannot be meaningfully applied to unseen payloads and is therefore excluded from this experiment. The evaluation thus compares \textsc{ForgeDAN} against GCG, AutoDAN-HGA, and human-designed DAN prompts. For consistency, all the adversarial components and configurations generated in §\ref{Jailbreaking Effectiveness} are reused directly, with only the malicious payload replaced, ensuring that the experiment isolates generalization capability rather than prompt construction.

The results in Table~\ref{tab:cross_sample} clearly demonstrate the superiority of \textsc{ForgeDAN}. Across all four target models, it achieves the highest ASR, ranging from 54.23\% on TranSpec-13B to 98.46\% on Gemma-2-9B. This substantially exceeds the performance of all baselines. AutoDAN-HGA shows moderate transferability with ASR between 14.42\% and 62.88\%, but its performance fluctuates significantly across models. Human-crafted DAN prompts achieve limited success (4.42–40.96\%), reflecting their lack of scalability. GCG performs particularly poorly in transfer settings, with success rates as low as 0.38\%, underscoring the brittleness of gradient-based suffix strategies. Notably, \textsc{ForgeDAN}’s advantage is especially pronounced on Gemma-2-9B and Qwen2.5-7B, where it surpasses the best baseline by over 40 percentage points.

Beyond overall superiority, several noteworthy insights become apparent. First, \textsc{ForgeDAN} maintains stable advantages across heterogeneous architectures, from open-source models (DeepSeek-V3, Gemma-2-9B, Qwen2.5-7B) to the proprietary TranSpec-13B, indicating that its evolutionary and semantically guided design is model-agnostic. Second, the relative margin between \textsc{ForgeDAN} and baselines tends to increase on larger or domain-specialized models, suggesting that its multi-strategy mutation and semantic fitness mechanisms are particularly effective in challenging transfer scenarios. Finally, the consistently high performance across all settings highlights its robustness, in contrast to the instability exhibited by AutoDAN-HGA and DAN.

\begin{table}[!tbp]
    \centering
    \caption{Comparison of Cross-Sample Generalization between \textsc{ForgeDAN} and Baselines}
    \label{tab:cross_sample}
    \scalebox{0.92}{
    \begin{tabular}{lcccc}
        \toprule
        \textbf{Target Models} & \textbf{\textsc{ForgeDAN}} & \textbf{GCG} & \textbf{AutoDAN-HGA} & \textbf{DAN} \\
        \midrule
        DeepSeek-V3   & \textbf{61.54\%} & 2.50\% & 14.42\% & 4.42\% \\
        Gemma-2-9B    & \textbf{98.46\%} & 0.38\% & 23.65\% & 23.65\% \\
        Qwen2.5-7B    & \textbf{87.12\%} & 5.38\% & 62.88\% & 40.58\% \\
        TranSpec-13B  & \textbf{54.23\%} & 16.39\% & 43.46\% & 40.96\% \\
        \bottomrule
    \end{tabular}
    }
\end{table}

\subsection{Real-World Applicability}
\label{Real-World Applicability}

In this section, we investigate the effectiveness of our approach when applied to real-world scenarios. While the previous experiments in §\ref{Jailbreaking Effectiveness} and §\ref{Generalization Analysis} evaluated effectiveness and generalization on the benchmark AdvBench dataset, this section leverages the proprietary real-world dataset introduced in §\ref{Dataset}, which was constructed from harmful chat records obtained from the operational logs of an anonymized AI enterprise. The same experimental setup as in §\ref{Generalization Analysis} is employed, with the only change being that the adversarial prompts are now evaluated on real-world payloads rather than benchmark samples. This design enables a direct assessment of practical utility in deployment contexts.  

The results in Table~\ref{tab:real_world} clearly show that \textsc{ForgeDAN} outperforms all baseline methods across the four target models. On Gemma-2-9B and Qwen2.5-7B, \textsc{ForgeDAN} achieves particularly strong performance, with ASR values of 100.00\% and 89.05\%, respectively. Even on the more challenging settings of DeepSeek-V3 and TranSpec-13B, it maintains success rates of 57.66\% and 56.20\%, substantially higher than the best-performing baselines. By comparison, GCG continues to exhibit weak performance (7.30–18.25\%), and while AutoDAN-HGA and DAN occasionally attain moderate success (up to 51.82\% on Qwen2.5-7B), their results remain far below those of \textsc{ForgeDAN}.  

Such observations reinforce the robustness of \textsc{ForgeDAN}. Unlike AutoDAN-HGA and DAN, whose performance varies considerably across models, \textsc{ForgeDAN} consistently ranks first, indicating stronger adaptability to heterogeneous real-world queries. These results confirm that the evolutionary design of \textsc{ForgeDAN}, grounded in semantic fitness and dual-dimensional verification, scales reliably to operational contexts. By consistently outperforming baselines across both general-purpose and domain-specific models, \textsc{ForgeDAN} demonstrates clear superiority in real-world jailbreak evaluations, directly addressing \textbf{\textit{RQ3}} and establishing itself as a practical and reliable framework for probing LLM safety in deployment environments.

\begin{table}[!tbp]
    \centering
    \caption{Comparison of Real-World Applicability between \textsc{ForgeDAN} and Baselines}
    \label{tab:real_world}
    \scalebox{0.92}{
    \begin{tabular}{lcccc}
        \toprule
        \textbf{Target Models} & \textbf{\textsc{ForgeDAN}} & \textbf{GCG} & \textbf{AutoDAN-HGA} & \textbf{DAN} \\
        \midrule
        DeepSeek-V3   & \textbf{57.66\%} & 11.68\% & 22.63\% & 18.98\% \\
        Gemma-2-9B    & \textbf{100.00\%} & 16.06\% & 19.71\% & 43.07\% \\
        Qwen2.5-7B    & \textbf{89.05\%} & 18.25\% & 49.64\% & 51.82\% \\
        TranSpec-13B  & \textbf{56.20\%} & 7.30\%  & 43.80\% & 46.72\% \\
        \bottomrule
    \end{tabular}}
\end{table}

\subsection{Ablation Study}
\label{Ablation Study}

The final research question (\textbf{\textit{RQ4}}) examines the relative contribution of each core component of \textsc{ForgeDAN}. To address this, we perform an ablation study by isolating the effects of its three main modules: the multi-strategic mutation mechanism, the semantic fitness measurement, and the dual-dimensional jailbreak judgment. In each ablation configuration, only one component is replaced with a simplified alternative, while the remaining components are preserved. The experimental setup—including 
seed malicious payload, target models, and ASR computation—follows that described in §\ref{Jailbreaking Effectiveness}, ensuring comparability with the full \textsc{ForgeDAN}. Specifically, the mutation module is restricted to synonym substitution, the fitness module is replaced with the cross-entropy metric introduced from AutoDAN-HGA, and the judgment module is downgraded to keyword matching.

\begin{figure}[!tbp]
\centering
\includegraphics[width=0.51\textwidth]{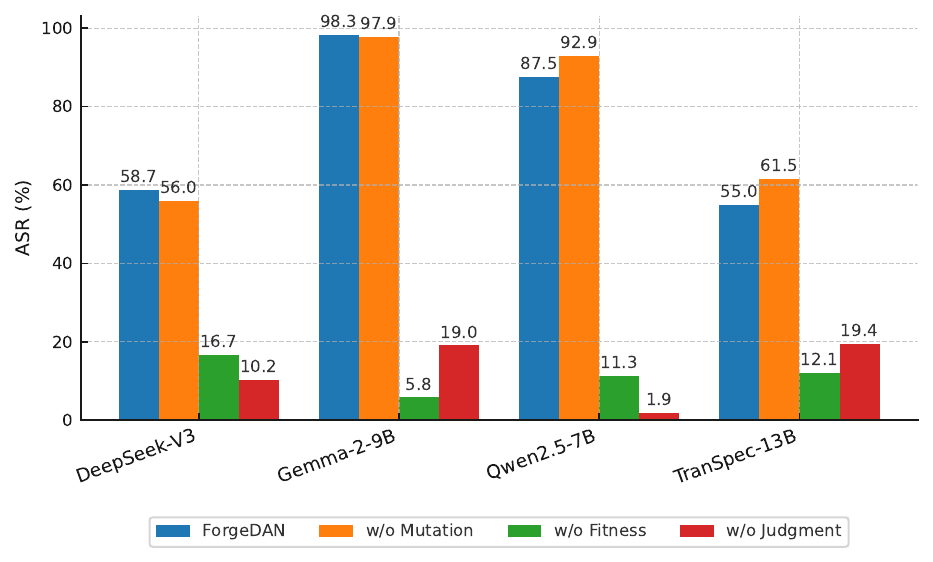}
\caption{Ablation study results.}
\label{fig:ablation_tudy}
\end{figure}

The results, demonstrated in Fig.~\ref{fig:ablation_tudy}, reveal distinct contributions of each module. Removing the multi-strategic mutation has a moderate impact: ASR remains relatively high across models (55.96\%–97.88\%), suggesting that synonym substitution alone can generate some successful adversarial prompts, although diversity and stability are reduced. In contrast, eliminating the semantic fitness measurement causes a dramatic decline in performance, with ASR dropping to as low as 5.77\% on Gemma-2-9B and 12.12\% on TranSpec-13B. This underscores the critical role of semantic-aware optimization in guiding the evolutionary process toward prompts that both preserve harmful intent and maximize attack success. Similarly, replacing the dual-dimensional jailbreak judgment with keyword matching sharply reduces ASR, with catastrophic failures on Qwen2.5-7B (87.50\% $\rightarrow$ 1.92\%) and DeepSeek-V3 (58.65\% $\rightarrow$ 10.19\%), highlighting the necessity of robust semantic discrimination to prevent misclassification and false positives.  

Further observations lead us to gain the following insights: when multi-strategic mutation module is reduced to simple synonym substitution, ASR remains relatively high. This shows that even a basic mutation strategy can generate effective adversarial prompts. However, the full, multi-strategic mechanism is still important for ensuring the \textbf{diversity and stability} of the generated prompts. Interestingly, in some cases, like with the Qwen2.5-7B model, the simplified version yielded a slightly higher ASR. This suggests that excessive or poorly-calibrated perturbations might sometimes introduce noise, hinting at opportunities for future refinement of adaptive mutation strategies.

Nevertheless, the semantic fitness module is a game-changer. Replacing it causes a dramatic collapse in performance. This stark decline underscores the module's crucial role in \textbf{guiding the evolutionary process}. It ensures that generated prompts maintain their malicious intent while becoming more effective at bypassing safety filters. Without this semantic-aware optimization, the attack simply becomes ineffective. Similarly, the dual-dimensional judgment module is indispensable. When it's downgraded to a simple keyword-matching mechanism, ASR drops sharply. This module's ability to perform \textbf{robust semantic discrimination} is essential for accurately identifying successful jailbreaks and preventing false positives. A simple keyword match isn't enough to capture the nuance of a model's response, leading to misclassification and an inability to correctly evaluate attack success.

In summary, the ablation results demonstrate that while each module provides distinct contributions, the synergy of multi-strategic mutation, semantic fitness measurement, and dual-dimensional judgment is essential for \textsc{ForgeDAN} to achieve reliable, transferable, and consistently high attack success across diverse models.

\section{Discussion}
\label{Discussion}

Building upon the experimental findings presented above, this section will discuss mitigation strategies for jailbreaking, computational cost trade-offs, and cross-model insights of jailbreak effectiveness that inform future research on jailbreak resistance.

\textbf{Jailbreaking Mitigation Strategies.} 
Mitigating jailbreak vulnerabilities in aligned LLMs requires a multi-layered defense that integrates training-time hardening, runtime safeguards, and robust refusal mechanisms. First, adversarial red-teaming samples should be systematically incorporated into safety-tuning pipelines. These adversarial prompts can serve as supervision data for SFT or as feedback trajectories in RLHF, enabling the model to learn consistent mappings from harmful or manipulative inputs to safe refusals rather than policy-violating outputs. Second, designing stronger runtime safety fences is essential. A promising approach is a dual-classifier architecture that distinguishes behavioral compliance detection (i.e., whether the model attempts to comply with an unsafe request) from semantic harmfulness detection (i.e., whether the generated content itself is unsafe). Third, enhancing the robustness and coverage of system prompts helps mitigate attacks that exploit static refusal patterns. Rather than depending on fixed templates, models should be trained to activate refusal behavior dynamically at any generation stage once potential risk is detected. Collectively, these measures—adversarially augmented alignment, layered runtime detection, and generalized refusal behaviors—constitute a practical and complementary defense-in-depth framework for improving LLM jailbreak resistance.

\textbf{Computational Efficiency and Cost Considerations.}
When evaluating automated jailbreak generation methods, it is essential to consider query costs and computational complexity across three stages: prompt generation, model inference, and result evaluation. Some baselines, such as manual DAN, GCG, and AutoDAN-HGA, incur negligible or no query costs. In contrast, PAIR and \textsc{ForgeDAN} involve all three stages—generation, inference, and judgment—thus incurring higher computational demands. PAIR’s iterative branching yields worst-case complexity of O(m·n), while \textsc{ForgeDAN} leverages parallelized evaluation, achieving O(m) per sample. Empirically, PAIR requires roughly an order of magnitude more time than \textsc{ForgeDAN} (tens of minutes versus minutes per jailbreak). Nevertheless, low-cost methods often trade efficiency for automation or generality: DAN lacks evolvability, GCG assumes white-box access, and AutoDAN-HGA’s keyword-based fitness limits diversity. Hence, evaluations should jointly consider ASR and computational overhead to strike a balanced view of effectiveness and efficiency.

\textbf{Cross-Model Variations in Jailbreaking Effectiveness.}
Differences in jailbreak success across target models reveal that vulnerability largely depends on model-specific alignment and safety mechanisms. Models with rigorous SFT and RLHF alignment or strongly detoxified data tend to refuse unsafe prompts more consistently, though over-alignment may cause excessive refusals of benign queries. Another key factor is the built-in safety architecture: some models enforce hard stops or fixed refusal templates, while others rely on adaptive filtering; the former resist generic attacks but are easier to bypass when their patterns are known. Refusal strategy robustness also varies—models with diverse, context-sensitive refusals exhibit higher resilience than those with predictable responses. However, because pretraining data, alignment strategies, and filtering details remain undisclosed for most open-source models, the causes of cross-model variability cannot yet be fully determined. Consequently, future evaluations should treat model heterogeneity as an inherent uncertainty and emphasize adaptive, model-agnostic red-teaming protocols.

\section{Conclusion}
\label{Conclusion}

This paper proposed \textsc{ForgeDAN}, an evolutionary jailbreak framework for LLMs. By integrating multi-strategy text mutations, semantic similarity–based fitness evaluation, and dual-dimensional jailbreak judgment, \textsc{ForgeDAN} overcomes the limitations of prior approaches that suffer from low diversity, shallow evaluation, and fragile detection.  

Extensive experiments on benchmark datasets and real-world scenarios show that \textsc{ForgeDAN} achieves higher attack success rates with greater naturalness and stealth compared to state-of-the-art baselines. These results highlight its value as both an automated red-teaming tool and a methodology for probing the safety boundaries of LLMs. Future work will explore co-evolutionary settings and extensions to multi-modal adversarial prompting.

\section*{Acknowledgment}
The authors would like to thank the anonymous reviewers for their valuable comments and suggestions. The research infrastructure and dataset of this work are supported by National Key Laboratory of Cognitive Intelligence and Anhui Provincial Laboratory of Safety Artificial Intelligence. We emphasize that our jailbreak framework is intended solely for red-teaming and improving LLM safety, not for malicious use.

\bibliography{myref}{}
\bibliographystyle{IEEEtran}

\end{document}